\documentstyle[12pt]{article}

\expandafter\chardef\csname pre amssym.def at\endcsname=\the\catcode`\@
\catcode`\@=11

\def\undefine#1{\let#1\undefined}
\def\newsymbol#1#2#3#4#5{\let\next@\relax
 \ifnum#2=\@ne\let\next@\msafam@\else
 \ifnum#2=\tw@\let\next@\msbfam@\fi\fi
 \mathchardef#1="#3\next@#4#5}
\def\mathhexbox@#1#2#3{\relax
 \ifmmode\mathpalette{}{\m@th\mathchar"#1#2#3}%
 \else\leavevmode\hbox{$\m@th\mathchar"#1#2#3$}\fi}
\def\hexnumber@#1{\ifcase#1 0\or 1\or 2\or 3\or 4\or 5\or 6\or 7\or 8\or
 9\or A\or B\or C\or D\or E\or F\fi}

\font\tenmsa=msam10
\font\sevenmsa=msam7
\font\fivemsa=msam5
\newfam\msafam
\textfont\msafam=\tenmsa
\scriptfont\msafam=\sevenmsa
\scriptscriptfont\msafam=\fivemsa
\edef\msafam@{\hexnumber@\msafam}
\mathchardef\dabar@"0\msafam@39
\def\dashrightarrow{\mathrel{\dabar@\dabar@\mathchar"0\msafam@4B}}
\def\dashleftarrow{\mathrel{\mathchar"0\msafam@4C\dabar@\dabar@}}

\def\ulcorner{\delimiter"4\msafam@70\msafam@70 }
\def\urcorner{\delimiter"5\msafam@71\msafam@71 }
\def\llcorner{\delimiter"4\msafam@78\msafam@78 }
\def\lrcorner{\delimiter"5\msafam@79\msafam@79 }
\def\yen{{\mathhexbox@\msafam@55 }}
\def\checkmark{{\mathhexbox@\msafam@58 }}
\def\circledR{{\mathhexbox@\msafam@72 }}
\def\maltese{{\mathhexbox@\msafam@7A }}

\font\tenmsb=msbm10
\font\sevenmsb=msbm7
\font\fivemsb=msbm5
\newfam\msbfam
\textfont\msbfam=\tenmsb
\scriptfont\msbfam=\sevenmsb
\scriptscriptfont\msbfam=\fivemsb
\edef\msbfam@{\hexnumber@\msbfam}
\def\Bbb#1{\fam\msbfam\relax#1}
\def\widehat#1{\setboxz@h{$\m@th#1$}%
 \ifdim\wdz@>\tw@ em\mathaccent"0\msbfam@5B{#1}%
 \else\mathaccent"0362{#1}\fi}
\font\teneufm=eufm10
\font\seveneufm=eufm7
\font\fiveeufm=eufm5
\newfam\eufmfam
\textfont\eufmfam=\teneufm
\scriptfont\eufmfam=\seveneufm
\scriptscriptfont\eufmfam=\fiveeufm

\catcode`\@=\csname pre amssym.def at\endcsname

\undefine\hbar
\newsymbol\ltimes 226E
\newsymbol\upharpoonright 1316
\newsymbol\hbar 207E
\newsymbol\square 1003
\newsymbol\blacksquare 1004

\newsymbol\hbar 207E
\def\Box{\hbox{\vrule height1ex\kern-0.4pt
\vbox to 1ex{\hrule width1ex\vfil\hrule width1ex}\kern-0.4pt\vrule height1ex}}
\newcommand{\sqr}[2]{{{\vcenter{\vbox{\hrule height.#2pt
\hbox{\vrule width.#2pt height#1pt \kern#1pt
\vrule width.#2pt}
\hrule height.#2pt}}}}}

\newcommand{\be}{\begin{equation}}

\newcommand{\ee}{\end{equation}}

\newcommand{\nn}{\nonumber}

\newcommand{\C}{{\Bbb C}}

\newcommand{\bib}{\bibitem}

\renewcommand{\H}{\mbox{$\cal H$}}

 \renewcommand{\ll}{\label}

\newcommand{\R}{{\Bbb R}}

\newcommand{\notp}{p \kern-.48em /}
\newcommand{\ci}{\cite}
\newcommand{\bea}{\begin{eqnarray}}
\newcommand{\eea}{\end{eqnarray}}

\newcommand{\half}{\mbox{\footnotesize $\frac{1}{2}$}}

\topmargin = - 0.5 cm
\begin{document}
\setlength{\baselineskip}{1.5\baselineskip}
\thispagestyle{empty}
\title{The St\"uckelberg-Kibble Model \\
as an Example of Quantized Symplectic Reduction}
\author{U.A.~ Wiedemann\thanks{supported by Bundesministerium
f\"ur Forschung und Technik (BMFT), F.R.\ Germany}\\
I. Institut f\"{u}r Theoretische Physik, Universit\"{a}t
Regensburg\\ Universit\"{a}tsstr. 31, D-93040 Regensburg,\\ F.R.\ Germany \\
\mbox{}\hfill \\ and \\
\mbox{}\hfill \\
 N.P.~ Landsman\thanks{Alexander von Humboldt Fellow and
S.E.R.C. Advanced Research Fellow}
 \\
 Department of Applied Mathematics and Theoretical Physics,\\
University of Cambridge, Silver Street,\\
Cambridge CB3 9EW, U.K.\thanks{Most of this work was done while the authors
were visiting the
II. Institut f\"{u}r Theoretische Physik, Universit\"{a}t Hamburg,
Luruper Chaussee 149, 22761 Hamburg, F.R. Germany}}
\maketitle
\begin{abstract}
Recently, it has been observed that a certain class of classical
theories with constraints can be quantized by a mathematical
procedure known as Rieffel induction.
After a short exposition of this idea, we apply the new
quantization theory to the St\"uckelberg-Kibble model.
We explicitly construct the
 physical state space ${\cal H}_{phys}$, which carries a massive
representation of the Poincar\'e group. The longitudinal
one-particle component arises from a particular Bogoliubov-transformation
of the five (unphysical) degrees of freedom one has started with.
Our discussion  exhibits the particular features of
the  proposed  	constrained  quantization theory in great clarity.
\end{abstract}
\newpage
\section{Introduction}
Classical gauge field theories may be defined by a set of fields ${\cal A}$,
subject to a set of constraints ${\cal B}$, which, in turn, generate
gauge transformations. The quantization of such theories is not a
unique procedure. Indeed, already in the two best-established
quantization methods very different technical setups are chosen.
On the one hand, one has the canonical operator formalism,
originating with Heisenberg and Pauli \ci{Heisenberg}, and now
well-adapted to handle non-abelian gauge theories \ci{Kugo},
whereas on the other hand Feynman's  path integral formalism
\ci{Feynman} allows the quantization of such theories through
the Faddeev-Popov procedure \ci{Faddeev}.
Both methods lead to identical perturbative expansions, but even
at a mathematically heuristic level their possible equivalence is
only known in perturbation theory.

It is certainly of general interest to have as many conceptually
and mathematically different quantization schemes as
possible, and to examine the particular features of each of them.
The hope of obtaining some hints on how to quantize gravity may
provide further motivation for  investigating  new quantization
schemes. Especially, the modern formulation of classical mechanics
in terms of symplectic manifolds and Poisson algebras
(see e.g. \ci{MaRa}) has suggested more refined quantization procedures,
such as  geometric quantization \ci{Woodhouse}, and strict deformation
quantization \ci{Rie91,NPLJGP}.

A particular feature of classical gauge theories that should somehow
be reflected in the quantization method is that the physical (reduced)
phase space  may be written as a so-called Marsden-Weinstein
quotient \ci{Arms,BFS}. It was shown in \ci{Landsman} that this classical
reduction procedure has a satisfactory quantum analogue in a procedure
from operator algebra theory known as Rieffel induction \ci{Rieffel}.
The way we apply this technique is mainly operator-theoretic, but a
certain aspect of the path integral formalism, viz.\ the integration over the
gauge group, will play a r\^{o}le as well.

This work discusses certain features of the  Rieffel induction procedure,
as applied to the quantization of constrained systems, which provides a
conceptually  and technically
new method for the quantization of certain gauge field theories.
 The method in question has already been  successfully applied
to certain finite-dimensional constrained systems \ci{Landsman},
as well as to free quantum electrodynamics \ci{Landswied,UAW}.

The present  work draws on these
results. Its aim is two-fold. Firstly, we would like to present
the strategy of this new quantization method in a form accessible
to a wider scientific community. Therefore, in Chapter 2,
we  briefly review the main line of argument, leading to the quantization
proposal. To keep our presentation reasonably short, we refer for some
of the technicalities to the aforementioned papers. Subsequently, in Chapter
3 we apply the new quantization scheme to the St\"uckelberg-Kibble model.
 This toy model has often been used
in the investigation of the Higgs mechanism and of spontaneous
symmetry breaking, see e.g.\ \ci{Morchio}.
 Here, we have chosen it since it  already shows many of the typical
complications of spontaneously broken  gauge theories without the need
to restrict oneself to a perturbative discussion.

 As we shall demonstrate explicitly for this model,
the Rieffel induction procedure provides a scheme
for the construction of the physical state space of a
constrained quantum theory, starting from a larger (unphysical) state
space on which the unconstrained theory is defined.
Our discussion will focus on the particular properties of this
  Rieffel-induced physical Hilbert space ${\cal H}_{phys}$.
Especially, we find that ${\cal H}_{phys}$ carries a trivial
representation of the gauge group and a massive representation of
the Poincar\'e group. Also, the positive spectrum condition turns out to be
satisfied. As an important by-product, we are able to trace back
how ``would-be Goldstone bosons rearrange to a massive,
longitudinal component'' in a theory exhibiting the Higgs mechanism.

The context of our work is modern symplectic geometry and reduction
theory on the classical side, and algebraic quantum field theory on
the quantum side. We only use the `soft' side of these
theories. Good recent introductions are \ci{MaRa,Haag,BW},
respectively.
\section{The quantization of gauge theories with Rieffel induction}
After presenting schematically the strategy which leads to Rieffel
induction in the quantization of theories with constraints, the
remainder of this section briefly specifies some notational and
technical prerequisites.

\subsection{Quantization of  Marsden-Weinstein
reduction}
The general symplectic reduction procedure, which is quantized by
Rieffel induction in its full generality, is described in \ci{Landsman}.
Here we are merely concerned with a special case, viz.\
Marsden-Weinstein reduction at the zero level of the moment map,
cf.\ \ci{MaRa,MW}. To introduce our notation, let us consider free
classical  electrodynamics. For the functional-analytic and other
details which are suppressed in what follows, we refer
the interested reader to \ci{Landswied}.

We start with the space $M$ of four-component  real-valued
weak solutions $A_{\mu}$   of the wave equation whose
Fourier-transformed Cauchy-data lie in
$ L^2(\R^3)\otimes \C^4 $. That is, $M =
{\lbrace{ A_{\mu} | {\Box}A_{\mu} = 0 }\rbrace}$. The imaginary part
\be
B(A,A') = 2{\rm Im}(A,A')_M
= -i \int {d^3{\bf p}\over (2\pi)^3} {\lbrack{A^{\mu}({\bf
p})\overline{A'}_{\mu}({\bf p}) - \overline{A}^{\mu}({\bf
p}) {A'}_{\mu}({\bf p})}\rbrack} \ll{sympl-form}
\ee
of the indefinite covariant scalar product $(*,*)_M$ turns $M$
into a symplectic space $(M,B)$, which is the phase space of the
unconstrained classical system. The set of constraints is given
by the gauge group $G$, which acts on $M$
via $A_{\mu} \to A_{\mu} + {\partial}_{\mu}g$, where
\be
G = {\lbrace{
g\in {\cal S}'(\R^4)\mid {\Box}g=0; dg \in M }\rbrace}.
\ee
Here, the space of distributions ${\cal S}'(\R^4)$ is the dual of
the usual Schwartz space of rapidly decreasing test functions. In
the present example, the reduced phase space $(M_c,B_c)$ of
the  corresponding constrained system may be obtained by a so-called
Marsden-Weinstein reduction \ci{MW}. This involves the moment map $J$
from $M$ into the dual of the Lie algebra of $G$. As $G$ is a vector
space, we may identify it with its Lie algebra, so we simply write
$J_g(A)$ for the value of $J(A)$ on $g\in G$.
Explicitly, the moment map turns out to be $J_g(A) = {\rm
Im}({\partial}g,A)_M$, cf. \ci{Landswied}. The preimage of its zero
level is
\be
J^{-1}(0) = {\lbrace{ A_{\mu}\in M \mid {\partial}_{\mu}A^{\mu} =
0}\rbrace}.
\ee
Then, $M_c$ is given by the Marsden-Weinstein quotient
\be
M_c = J^{-1}(0) / G,
\ee
and $B_c$ inherits its structure from $B$. It is easy to see that
$(M_c,B_c)$ defined this way indeed describes the physical degrees
of freedom of free electrodynamics: picking $J^{-1}(0)$ fixes the
gauge (thus imposing the Gauss law constraint, which on elements of
$M$ becomes the Lorentz gauge condition), and quotienting by $G$
removes the gauge degeneracy of the symplectic form $B$ with respect
to the action of $G$ on $J^{-1}(0)$.

In principle, there are two possibilities to quantize a reduced phase
space $(M_c,B_c)$. Either, we directly quantize  the Marsden-Weinstein
reduced (i.e. constrained) classical system $(M_c,B_c)$, or we quantize
the unconstrained classical system $(M,B)$ together with the set of
constraints. In the latter case, a scheme has to be found which
imposes constraints on the unconstrained quantized theory,
thereby providing a quantum analogue of the classical
Marsden-Weinstein reduction. Examples of such schemes are
the Dirac or the BRST method. According to the proposal of \ci{Landsman},
the so-called Rieffel induction procedure of operator algebra
theory \ci{Rieffel} (which we explain below) provides a rival scheme,
which in all examples studied so far works as well as, or better than
the methods mentioned above.

More precisely, let us consider schematically a quantization prescription
$Q_{\hbar}$ which relates the symplectic space $(M,B)$ (or rather the
Poisson algebra of functions on it) to some algebra of field  operators
on a Hilbert space ${\cal A}$, $G$ to some algebra ${\cal B}$ generated
by $G$, and  $(M_c,B_c)$ to some (a priori unknown) algebra of
observables (in the sense of gauge-invariant operators) ${\cal A}_{obs}$.
Then, according to our quantization proposal, the following diagram
commutes:
\be
\begin{array}{ccc}
(M,B) ; G &
   \stackrel{Q_{\hbar}}{\longrightarrow}& {\cal A}; {\cal B}\\
\vcenter{
  \llap{$\scriptstyle{\rm Marsden-Weinstein\,\, Reduction}$}}
\Big\downarrow&
         &\Big\downarrow\vcenter{
           \rlap{$\scriptstyle{\rm Rieffel\,\, Induction}$}}\\
(M_c,B_c) &
   \stackrel{Q_{\hbar}}{\longrightarrow}& {\cal A}_{obs}
\end{array} \ll{diag1}
\ee
Our program in this paper is to specify the entries of this
diagram for the St\"uckelberg-Kibble model. To this end,
  we briefly recall how, for a linear field theory, a symplectic space
$(M,B)$ can be related to a field algebra ${\cal A}$ of canonical commutation
relations, and we explain how Rieffel induction allows one to construct
new Hilbert spaces for quantum field theories, thereby eventually specifying
${\cal A}_{obs}$.
\subsection{Weyl algebras of canonical commutation relations}
The general theory behind this subsection is explained in great detail
and rigour in, e.g., \ci{BW}, and the application to electromagnetism
is from \ci{CGH}. We merely mention some of the main points.

For ${\phi}, {\phi}' \in M$, the operators $W({\phi})$, $W({\phi}')$,
satisfying the Weyl form of the canonical commutation relation (CCR)
\be
W({\phi})W({\phi}') = W({\phi}+{\phi}')
e^{{-i\over 2}B({\phi},{\phi}')}, \ll{ccr-qed-weyl}
\ee
 specify a field algebra with $C^*$-structure which we denote by
${\cal A}(M,B)$. In most cases, one is primarily interested in
the properties of the operator vector potential $A_{\mu}$, for which
we use the same notation as for its classical counterpart, as no
confusion will arise. The $A_{\mu}$ satisfy the canonical commutation
relations
\be
[A_{\mu}(x), A_{\nu}(y)] = -ig_{\mu\nu} D(x-y),\ll{ccr-qed-vect}
\ee
 where $D$ denotes the commutator function satisfying $\Box{D}=0$, with
initial conditions $D({\bf x},0)=0$,
${\partial\over{\partial t}} D({\bf x},t){\mid}_{t=0} =
-{\delta}^{(3)}({\bf x})$. To see the connection between
(\ref{ccr-qed-weyl}) and (\ref{ccr-qed-vect}), we consider the vector
potential $A(f) = \int d^4x A_{\mu}(x)f^{\mu}(x)$, smeared with real
test functions $f$. Now, (\ref{ccr-qed-vect})
reads $[A(f),A(g)]=i\sigma(f,g)$, where
$\sigma(f,g) = -\int d^4x d^4y D(x-y)f^{\mu}(x)g_{\mu}(y)$.
Formally, this allows for the introduction of the operators
$U(f)=e^{[iA(f)]}$ which according to the
Baker-Campbell-Haussdorff formula satisfy the Weyl form of the
canonical commutation relations
$U(f)U(g) = U(f+g) e^{[-{i\over 2}\sigma(f,g)]}$.
Here, however, $U(f)$ and $U(f')$ have the same commutation relations
as long as $\int d^4x D(x-y) (f^{\mu}(x) - {f'}^{\mu}(x)) = 0$
for almost all $y$. To remove this degeneracy and
to obtain a one-to-one correspondence between Weyl operators and test
functions, one uses the map $f\to\phi$, defined by the
convolution ${\phi}_{\mu} = D*f_{\mu}$. Then, the space $M$ of
solutions of the wave equation $\Box{\phi}_{\mu}=0$,
$ {\phi}_{\mu}({\bf x},t) = {1\over {{(2\pi)}^3}} \int {d^3{\bf k}\over
2k_0} [{\phi}_{\mu}({\bf k})e^{-ikx}$ $+
\overline{{\phi}_{\mu}}({\bf k})e^{ikx}]$, is\footnote{
Our notation does not distinguish between functions $\phi$ and their Fourier
transforms, since no confusion should arise.}
\be
M = \overline{\lbrace{ \phi = D*f}\rbrace} =
L^2({\R}^3)\otimes{{\C}\,}^4. \ll{M-qed}
\ee
 Now, the operators $W(\phi) = U(f)$, $\phi\in M$
satisfy (\ref{ccr-qed-weyl}) with symplectic form $B$ induced
by $\sigma$ and given in (\ref{sympl-form}).

Having established the connection between Weyl operators and
vector potentials, we can
introduce formal annihilation and creation operators
$a_{\mu}$, $a_{\mu}^*$. E.g. for the free electromagnetic field,
$ A_{\mu}(x) = \int {d^3{\bf k}\over (2\pi)^32k_0}\,
{\lbrack{e^{-ikx} a_{\mu}({\bf k}) + e^{ikx} a_{\mu}^*({\bf
k})}\rbrack}{\mid}_{k_0={\bf k}}$,
\be
iA(f) = \int {d^3{\bf k}\over {(2\pi)}^32k_0} {\lbrack{ a_{\mu}({\bf
k})\overline{{\phi}_{\mu}}({\bf k}) - a_{\mu}^*({\bf k})
{\phi}_{\mu}({\bf k})}\rbrack} =: a_{\mu}({\phi}^{\mu}) -
a_{\mu}({\phi}^{\mu})^*. \ll{crean}
\ee
Clearly, in terms of the annihilation and creation operators, the
Weyl operators read
$ W({\phi}^{\mu}) = \exp{\lbrack{a_{\mu}({\phi}^{\mu}) -
a_{\mu}({\phi}^{\mu})^*}\rbrack}$, where
${\lbrack{a_{\mu}({\phi}^{\mu}), a_{\nu}({{\phi}'}^{\nu})^*}\rbrack}
= ({\phi}',{\phi})_M$, $(.,.)_M$ denoting the indefinite Minkowski inner
product. Heuristically, one has
\be
{d\over d\lambda}W(\lambda{\phi})|_{\lambda=0} = iA(f).\ll{deriv-W}
\ee
 It is well-known that this derivative does not exist in the
operator norm but with respect to regular representations only,
and thereby the $a_{\mu}$, $a_{\mu}^{*}$ only exist in such
representations, too. Nevertheless, in what follows we shall adopt
the formal expressions (\ref{crean}) and (\ref{deriv-W}), even when
no explicit reference to a particular representation is made.

As a final preparatory step, we point out that subalgebras
of ${\cal A}(M,B)$ can be specified by selecting subspaces
of $M$. In particular, for free QED,
\bea
N &=& {\lbrace{ {\phi}_{\mu}\in M | k^{\mu}{\phi}_{\mu}({\bf k}) = 0}
      \rbrace} = {\lbrace{ {\phi}_{\mu}\in M | {\partial}^{\mu}
      {\phi}_{\mu}(x) = 0}\rbrace}, \nn \\
T &=& {\lbrace{ {\phi}_{\mu}\in M | {\phi}_{\mu}({\bf k}) =
      ik^{\mu}g({\bf k})}\rbrace} =
      {\lbrace{ {\phi}_{\mu}\in M | {\phi}_{\mu}(x) =
      {\partial}^{\mu}g(x), \Box g(x) = 0}\rbrace} \nn \\ \ll{subspace}
\eea
define subalgebras ${\cal A}(N,B)$, ${\cal A}(T,B)$ of
${\cal A}(M,B)$. Note that $T\subset N$, so that
${\cal A}(T,B)$ $\subset {\cal A}(N,B)$. These
subalgebras are Poincar\'e-invariant, as may be seen by
recalling that the action of elements $(\Lambda, a)$ of the
Poincar\'e group ${\cal P}$ on ${\cal A}(M,B)$ is defined via
the algebraic automorphism ${\alpha}_{(\Lambda, a)}$,
\be
{\alpha}_{(\Lambda, a)}(W({\phi}^{\mu})) =
W({\gamma}_{(\Lambda, a)}({\phi}^{\mu})) \qquad\mbox{with}\qquad
({\gamma}_{(\Lambda, a)}({\phi}^{\mu}))(x) =
{\Lambda}^{\mu}_{\nu}{\phi}^{\nu}({\Lambda}^{-1}(x-a)).\ll{Poincare}
\ee

\subsection{Rieffel induction}
This subsection gives a quick `review by example' of some parts of
the theory developed in \ci{Landsman} and \ci{Landswied}.

In physics, induction methods are mainly known from Wigner's
classification and construction of all irreducible unitary
representations of the Poincar\'e group $P$. In general, the method of
induced representations of (locally compact) groups allows one to
construct a   representation of the complete group from
a representation of a subgroup, cf.\ e.g.\ \ci{Barut}.

Also, in the theory of operator algebras (particularly $C^*$-algebras)
a method exists for constructing a representation of an algebra,
given a representation of  some other algebra \ci{Rieffel}.
The latter  is not necessarily a subalgebra of the former;
instead, the two algebras need to be connected by a bimodule
with certain additional properties. Whatever the technical details,
the main idea is that the representation one induces from should be
straightforward, and yet capable of producing an appropriate
representation of the algebra one is really interested in.
This idea will be fully realized in our context, for the second
algebra will be the algebra generated by the gauge group, and the
representation induced from is the trivial one. With a suitable
choice of bimodule, the induced representation of the algebra of
observables comes out to be the vacuum representation
on a Fock space of physical photon states.

To facilitate our presentation, we proceed by example, abstracting
general features afterwards. For free QED,  in the diagram (\ref{diag1})
we choose the field algebra ${\cal A} = {\cal A}(M,B)$ and the
`algebra of constraints' ${\cal B} = {\cal A}(T,B)$ (cf.\ the previous
subsection), where the choice of ${\cal B}$ is motivated by observing
that the gauge group $G$ equals $T$,
cf. (\ref{subspace}).\footnote{ For simplicity, we  here ignore
some mathematical difficulties in defining algebras ${\cal B}$
for groups $G$ which are not locally compact. This greatly
simplifies our presentation. For more details, we refer to
\ci{Landswied,UAW}.}
Also, we introduce the `algebra of weak observables' ${\cal A}_c :=
{\cal A}(N,B)$, which is the largest subalgebra of
${\cal A}={\cal A}(M,B)$ commuting with ${\cal B} = {\cal A}(T,B)$.

The Rieffel induction procedure will produce a representation of
${\cal A}_c$ induced from a representation of $\cal B$. To this end,
we need a bimodule for ${\cal A}_c$ and $\cal B$, that is, a
linear space on which ${\cal A}_c$ acts from the left, and $\cal B$
acts from the right (that is, in an anti-representation), so that these
two actions commute. In the case at hand, ${\cal A}_c$ and ${\cal B}$,
which is abelian, are each other's commutant in the field algebra $\cal
A$, so that  a representation of $\cal A$ on a Hilbert space $\cal H$
automatically defines such a bimodule. Finally, we need a representation
of $\cal B$ to induce from. This is the trivial one,
defined on the Hilbert space  ${\cal H}_{tr}=\C$. Schematically,
\be
{\cal A}_c \longrightarrow {\cal H} \longleftarrow {\cal B}
\longrightarrow {\cal H}_{tr}. \ll{diag2}
\ee
 The restriction of the action $\pi$ on $\H$  of $\cal A$ to its
subalgebra $\cal B$ defines a representation $U$ of the gauge group,
that is, one has $U(\phi)=\pi(W(\phi))$.

 As will be discussed in more detail below, this setup allows the
construction of a positive semidefinite sesquilinear form $(.,.)_0$
on $L\otimes {\cal H}_{tr}$, where $L$ is a suitable dense subspace
of $L$. In the present case, this form is given by
\be
(\psi\otimes v,\varphi \otimes w)_0 = v\overline{w}
\int_G [{\cal D}\phi] (U(\phi) \psi,\varphi). \ll{0-prod}
\ee
Here $[{\cal D}\phi]$ denotes the non-existent `Lebesgue' measure on
the gauge group $G$. The point is, however, that this flat `measure'
combines with a factor in the integrand to define a mathematically
well-defined  path integral (cylindrical) measure on  $G$ \ci{Landswied}.
Furthermore, $\psi, \varphi $ are in $ {\cal H}$,
$v,w \in {\cal H}_{tr} = {\C}$, and $(.,.)$ is the inner product on $\H$.

Irrespective of the explicit form of $(.,.)_0$, the induced
physical Hilbert space is then defined as the completion of the
quotient of $L\otimes {\cal H}_{tr}$ by the null space of $(.,.)_0$,
i.e.,
\be
{\cal H}_{phys} = \overline{(L\otimes {\cal H}_{tr}) / {\cal N}}, \ll{quotient}
\ee
where ${\cal N} \subset L\otimes {\cal H}_{tr}$ is the subset of
vectors with vanishing $(.,.)_0$ norm. The collection of vectors in
${\cal H}_{phys}$ of the form $\psi\tilde{\otimes}v$, defined as the
image of $\psi\otimes v\in L\otimes {\cal H}_{tr}$ under the
quotient projection from $L\otimes {\cal H}_{tr}$ to ${\cal H}_{phys}$,
are clearly dense in ${\cal H}_{phys}$. The action of
elements $A$ of ${\cal A}_c$ on ${\cal H}_{phys}$ is then given on this
dense set by
${\pi}_{phys}(A)\psi\tilde{\otimes}v = ({\pi}(A)\psi)\tilde{\otimes}v$.
Under appropriate continuity conditions \ci{Rieffel,Landswied} this
action may be extended to all of ${\cal H}_{tr}$.

The reader should note that ${\cal H}_{phys}$ satisfies an essential
requirement of a non-degenerate physical Hilbert space:
the gauge degeneracy of elements of ${\cal A}_c$ is removed in
${\pi}_{phys}({\cal A}_c)$. To see this, choose an arbitrary element
$W(\phi) \in {\cal A}_c$. From equation (\ref{0-prod}), it is obvious
that for ${\phi}_t \in T$ (which, we recall, coincides with the gauge
group $G$), ${\pi}_{phys}(W(\phi))\psi\tilde{\otimes}v =
{\pi}_{phys}(W(\phi + {\phi}_t))\psi\tilde{\otimes}v$
for all vectors $\psi\tilde{\otimes}v \in {\cal H}_{phys}$. Hence,
${\pi}_{phys}(W(\phi)) = {\pi}_{phys}(W(\phi + {\phi}_t))$. This
removal of the gauge degeneracy of ${\cal A}_c$ is independent of the
choice of ${\cal H}$, and hence we indentify ${\pi}_{phys}({\cal A}_c)$
with the representation-independent algebra of observables ${\cal A}_{obs}$,
cf.\ (\ref{diag1}).

Let us now turn to the abstract setting which has led to the
$(.,.)_0$-inner product (\ref{0-prod}). As stated, the aim of the
Rieffel induction procedure is to obtain a representation
${\pi}_{phys}$ of ${\cal A}_c$ induced from a representation of
${\cal B}$ on some Hilbert space ${\cal H}_{\chi}$. Our example,
and all similar examples involving gauge theories, have the special
feature that  ${\cal H}_{\chi}={\cal H}_{tr}={\C}$, that
is, one induces from the trivial representation of the gauge group.
This will imply that the algebra of constraints ${\cal B}$ is
represented trivially on the induced space ${\cal H}_{phys}$.
Technically, the construction of ${\pi}_{phys}$ proceeds
according to the following three step method.
\begin{enumerate}
\item
Given a bimodule $L$ for ${\cal A}_c$ and $\cal B$, a ${\cal B}$-valued
scalar product ${\langle{.,.}\rangle}_{\cal B}$ has to be found on $L$,
that is, for $\psi, \varphi \in L \subset {\cal H}$,
${\langle{\psi,\varphi}\rangle}_{\cal B} \in {\cal B}$,\footnote{
Mathematically, ${\langle{.,.}\rangle}_{\cal B}$ is a so-called
rigging map which has to satisfy the
following conditions for all $\psi$, $\varphi\in L$ \ci{Rieffel}:
\begin{enumerate}
\item
${\langle{\lambda\psi,\mu\varphi}\rangle}_{\cal B} =
\overline{\lambda}\mu{\langle{\psi,\varphi}\rangle}_{\cal B}$
$\hbox{   }$ for all ${\lambda},\mu\in{\C}$;
\item
${\langle{\psi,\varphi}\rangle}_{\cal B}^* =
{\langle{\varphi,\psi}\rangle}_{\cal B}$ (where the $\mbox{}^*$ denotes
the hermitian conjugate in $\cal B$);
\item
${\langle{\psi,\varphi
B}\rangle}_{\cal B} = {\langle{\psi,\varphi}\rangle}_{\cal B}B$
$\hbox{   }$ for all $B\in {\cal B}$ (on the left-hand side, $B$
 acts in the given right-representation on the bimodule $L$,
whereas on the right-hand side $B$ acts by multiplication
in the algebra $\cal B$);
\item
${\langle{A\psi,\varphi}\rangle}_{\cal B} =
{\langle{\psi,A^*\varphi}\rangle}_{\cal B}$ $\hbox{   }$ for all $A\in
{\cal A}$.
\item
${\langle{A\psi,A\psi}\rangle} \leq {\parallel{A}\parallel}^2
{\langle{\psi,\psi}\rangle}$ $\hbox{   }$ for all $\psi \in L$, $A\in
{\cal A}$.
\end{enumerate}}
\item
Given such an operator-valued scalar product, the  tensor product
$L\otimes {\cal H}_{\chi}$ is equipped with a sesquilinear form $(.,.)_0$,
\be
(\psi\otimes v,\varphi\otimes w)_0 := ({\pi}_{\chi}({\langle{\varphi,
\psi}\rangle}_{\cal B})v,w)_{\chi}.\ll{rigged}
\ee
Crucially, this form is positive-semidefinite if the postivity
condition \\ $\pi_{\chi}
({\langle{\psi,\psi}\rangle}_{\cal B})\geq 0$ for all $\psi\in L$
is satisfied, which is the case in all our examples.
\item
The subspace ${\cal N} \subset L\otimes {\cal H}_{tr}$ of vectors
with vanishing $(.,.)_0$-norm is determined and the physical
Hilbert space is defined as in
(\ref{quotient}).
\end{enumerate}
The most difficult part of this procedure is to find
${\langle{.,.}\rangle}_{\cal B}$. Here, one is guided by
mathematical examples \ci{Landsman}.
One may consider e.g. ${\cal B} = C^*(G)$, the $C^*$-group
algebra of a locally compact group $G$  (cf.\ \ci{BW}; this is
essentially the convolution algebra on the group w.r.t.\ the Haar
measure). Then,  it can be shown that a rigging map
${\langle{.,.}\rangle}_{\cal B}$ is defined as follows:
${\langle{\psi,\varphi}\rangle}_{\cal B}$
has to be some element of $C^*(G)$, i.e., a function on the group,
and we prescribe that the value of this function at $g\in G$ is given by
 ${\langle{\psi,\varphi}\rangle}_{\cal B} (g) = (U(g)\varphi,\psi)$,
where $U$ is a continuous unitary representation of $G$ on  ${\cal H}$,
commuting with ${\pi}({\cal A}_c)$, $x\in G$. Inducing from the trivial
representation ${\cal H}_{tr} = {\C}$, one obtains\footnote{
In what follows, we use the shorthand $(\psi,\varphi)_0$ for
$(\psi\otimes v,\varphi\otimes w)_0$, since $v,w \in {\C}$ are
complex numbers which can be absorbed in the definition of
$\psi$ and $\varphi$.}
\be
(\psi,\varphi)_0 = \int_G dx (U(x)\psi,\varphi), \ll{0-prod-G}
\ee
of which (\ref{0-prod}) is a special case, at least in a heuristic sense.

In what follows, we shall take a suitable
generalization of (\ref{0-prod-G})  as our starting point,
thereby obviating the need for a discussion of the explicit form
and a verification of the mathematical properties of
${\langle{.,.}\rangle}_{\cal B}$. In fact, our presentation of the
Rieffel induction procedure for quantum field theories has been
slightly oversimplified with respect to this point. While the
existence of a so-called `rigged' inner product $(.,.)_0$, defined
in (\ref{rigged}), is always sufficient for the quantization  proposal to
apply, it is not always possible to derive it from a mathematically
well-defined rigging map ${\langle{.,.}\rangle}_{\cal B}$. We refer
to \ci{Landswied} for a discussion of the technical points involved.

To sum up: In this chapter, we have seen that Rieffel induction
provides a well-defined scheme for the construction of a physical
Hilbert space ${\cal H}_{phys}$, on which gauge transformations
act trivially. In the corresponding algebra of observables
${\pi}_{phys}({\cal A}_c)$, all gauge degeneracies are removed,
i.e., Rieffel induction is a method to impose constraints on
quantum field theories. The physical Hilbert space ${\cal H}_{phys}$
is obtained by forming the quotient of a larger Hilbert space
$L\otimes {\cal H}_{tr}$ with respect to a null space.

This is somehwat reminiscent of the BRST (or, in case of QED,
the Gupta-Bleuler) procedure, with the major difference that with
Rieffel induction no negative-norm subspace exists, obviating the
need to select a physical subspace of $\cal H$. Also, certain
functional-analytic problems that appear in the BRST as well as
in the Dirac method are absent with our present
techniques \ci{Landsman,Landswied}. By definition of the inner
product on the physcial Hilbert space ${\cal H}_{phys}$, calculations
of correlation functions of operators in ${\cal A}_c$ (as represented
on ${\cal H}_{phys}$) may be performed in $L\otimes {\cal H}_{tr}$,
\ci{UAW}.
\section{Application to the St\"uckelberg-Kibble model}
In this chapter, we specify (\ref{diag2}) and
(\ref{0-prod}) for the St\"uckelberg-Kibble model, thereby
constructing a physical Hilbert space ${\cal H}_{phys}$ for
this model.
The St\"uckelberg-Kibble model is an abelian Higgs model with the
modulus $\eta$ of the scalar field $\phi(x) = \eta(x)
e^{\varphi(x)}$ frozen to unity, $\eta(x) =1$. It is given by the
Lagrangian
\be {\cal L}= -{1\over 4} F_{\mu\nu}F^{\mu\nu} - {1\over 2} \left(
{\partial}_{\mu}\varphi + eA_{\mu}\right)\left(
{\partial}^{\mu}\varphi + eA^{\mu}\right).
\ee
Despite its linearity, this model has non-trivial features, and has
been used as testing ground for investigations of the Higgs mechanism
before \ci{Morchio}. Its equations of motion can be written in terms
of a gauge-invariant current $j^{\mu} =
{\partial}^{\mu}\varphi + e A^{\mu}$, satisfying
\be
\left( \Box + e^2\right) j^{\mu} = 0 \qquad \hbox{;}\qquad
{\partial}_{\mu}j^{\mu} = 0. \ll{j-equ}
\ee
In fact, this is nothing but the Proca equation \ci{Itzykson} of
a massive gauge-invariant vector field. To make this model amenable
to treatment by symplectic reduction and quantum induction methods, we
now make a move that is analogous to rewriting the Maxwell equation
for $A_{\mu}$ as a massless Klein-Gordon equation plus a subsidiary
Lorentz condition. Thus we pass back to the gauge-dependent fields
$A_{\mu}$ and $\varphi$, and choose what is essentially the 't Hooft
gauge as the subsidiary condition:
\be {\partial}_{\mu} A^{\mu} = e\varphi. \ll{thg}
\ee
With this constraint, the equations of motion read
\be
\left( \Box + e^2\right) A^{\mu} = 0 \qquad \hbox{,} \qquad
\left( \Box + e^2\right) \varphi = 0, \ll{g-equ}
\ee
and the gauge group $G={\lbrace{g| \left( \Box + e^2\right)g = 0}\rbrace}$
acts on $A_{\mu}$, $\varphi$ via
\be
A_{\mu} \to A_{\mu} + {\partial}_{\mu}g \qquad\mbox{ , }\qquad
\varphi \to \varphi - eg. \ll{g-trafo}
\ee
\subsection{Marsden-Weinstein reduction for the St\"uckelberg-Kibble
model}
A mathematically rigorous treatment of the following material, in the
style of \ci{Landswied}, is possible, but we leave the details to the
interested reader; instead, readability commands us to give
somewhat loose formulations.

 Our investigation of the St\"uckelberg-Kibble model starts from
the symplectic space $(M_{sk},B_{sk})$, defined by
\bea
&& M_{sk} = {\lbrace{ ( A_{\mu},{\varphi}) \mid  A_{\mu}\in
L^2({\R}^3)\otimes {{\C}\,}^4, {\varphi}\in L^2({\R}^3); \left( \Box +
e^2\right) A_{\mu}= \left( \Box + e^2\right){\varphi}=0
}\rbrace},\nn \\
&&B_{sk}( A_{\mu},{\varphi};{A'}_{\mu},{\varphi}') =
2{\rm Im}( A_{\mu},{A'}_{\mu})_M - 2{\rm Im}({\varphi},{\varphi}').
\ll{sk-sympl}
\eea
The gauge group $G$ acts on this space by the gauge transformation
(\ref{g-trafo}). This action is strongly Hamiltonian, and hence,
in particular, it is symplectic. We evidently may identify the gauge
group with the following subspace of  $M_{sk}$
\be
T_{sk} = {\lbrace{ ( A_{\mu},{\varphi})  \in M_{sk} \mid  A_{\mu}=
{\partial}_{\mu}g,  {\varphi}= -eg; g\in L^2({\R}^3);
\left( \Box + e^2\right) g = 0}\rbrace}.\ll{tsk}
\ee
{}From this, the Marsden-Weinstein reduced space $(M_{c,sk},B_{c,sk})$
is easily calculated. With similar notation as in subsection 2.1,
the moment map reads
\be
J_g( A_{\mu},\varphi) =
2{\rm Im}({\partial}_{\mu}g,A_{\mu})_M
- 2{\rm Im}(-eg,{\varphi}),
\ee
which leads to
\be
J^{-1}(0) = \{( A_{\mu},{\varphi}) \in M_{sk}
| {\partial}_{\mu} A^{\mu} = e\varphi\}. \ll{jmoo}
\ee
Hence in view of (\ref{thg}) the Marsden-Weinstein quotient reads
\be
M_{c,sk} = J^{-1}(0)/G = {\lbrace{ j_{\mu} \in
L^2({\R}^3)\otimes {{\C}\,}^4| \left( \Box + e^2\right)j_{\mu}=0 ;
{\partial}_{\mu}j^{\mu} = 0}\rbrace}.
\ee
The symplectic form $B_{c,sk}$ on $M_{c,sk}$ inherits its structure
from $B_{sk}$, and is given by
\be
B_{c,sk}(j,j') = {2\over e^2} {\rm Im}(j_{\mu}, {{j}'}_{\mu})_M.
\ee
Clearly, $(M_{c,sk},B_{c,sk})$ is the phase space of a massive vector
boson, which indeed represents the physical degrees
of freedom of the St\"uckelberg-Kibble model. This
completely specifies the left-hand side  of the diagram (\ref{diag1}).
\subsection{Rieffel induction for the St\"uckelberg-Kibble model}
\subsubsection{Construction of the field algebra}
Consider the canonical commutation relations of the operator fields
$A_{\mu}$ and $\varphi$ (denoted by the same
symbol as their classical counterparts):
\bea
[{\varphi}(x),{\varphi}(y)] &=& i{\triangle}(x-y), \nn \\
{[ A_{\mu}(x), A_{\nu}(y)]} &=&
-ig_{\mu\nu}{\triangle}(x-y), \ll{ccrsk}
\eea
where the commutator function ${\triangle}$ satisfies
$(\Box + e^2){\triangle}(x) =0$ with initial conditions
${\triangle}({\bf x},0) =0$, ${{\partial}\over {\partial}t}
{\triangle}({\bf x},t){\mid}_{t=0} = -{\delta}^{(3)}({\bf x})$.
In analogy with our discussion of free QED, we specify
the formal connection between the fields $A_{\mu}$, $\varphi$
and the corresponding Weyl operators,
$
W({\phi}_{\mu},{\phi}) = e^{iA_{\mu}(f^{\mu}) +
i\varphi(f)},$
where ${\phi}_{\mu} = \triangle * f_{\mu}$, ${\phi} = \triangle * f$.
Here, either as a consequence of (\ref{ccrsk}), or imposed axiomatically,
the operators $W({\phi}_{\mu},{\phi})$, $W({{\phi}'}_{\mu},{\phi}')$
satisfy the Weyl form of the canonical commutation relations
\be
W({\phi}_{\mu},{\phi})W({\phi}_{\mu}',{\phi}')
= W({\phi}_{\mu} + {\phi}_{\mu}',{\phi} + {\phi}') e^{-{i\over
2}B_{sk}({\phi}_{\mu},{\phi};{{\phi}'}_{\mu},{\phi}')}.
\ee
The field algebra of the model is then defined as the Weyl algebra
${\cal A}(M_{sk},B_{sk})$ generated by the $W$'s subject to these
commutation relations (cf.\ \ci{BW}).

Now, we want to construct the quantum counterpart of Marsden-Weinstein
reduction, i.e., we want to complete the right hand side of
the diagram (\ref{diag1}). Therefore, we invoke the
quantization prescription for symplectic spaces as discussed in
Chapter 2. This leads to the field algebra
${\cal A}\equiv  {\cal A}(M_{sk},B_{sk})$ defined by (\ref{sk-sympl}).
Also, in analogy with our discussion in Chapter 2, we choose the algebra of
constraints ${\cal B}={\cal A}(T_{sk},B_{sk})$; once again, the motivation
for this is that it is the ($C^*$) algebra generated by the gauge group.
Consequently, the algebra of weak observables, which by
definition is the largest subalgebra of ${\cal A}(M_{sk},B_{sk})$
commuting with ${\cal B}(T_{sk},B_{sk})$, is given by ${\cal
A}_c={\cal A}(N_{sk},B_{sk})$, where
\be
 N_{sk} = {\lbrace{ ({\phi}_{\mu},{\phi}) \mid
{\partial}^{\mu}{\phi}_{\mu}= e{\phi} }\rbrace} \subset M_{sk}; \ll{nsk}
\ee
compare this with (\ref{jmoo}).
The subspaces $N_{sk}$ and $T_{sk}\subset N_{sk}$ of $M_{sk}$ are
invariant under the action of symplectic transformations
${\gamma}_{\Lambda,a}$ associated with elements $(\Lambda,a)$
of the Poincar\'e group ${\cal P}$,
$({\gamma}_{\Lambda,a}({\phi}_{\mu},{\phi}))(x) :=
({\Lambda}^{\nu}_{\mu}{\phi}_{\nu},{\phi})({\Lambda}^{-1}(x-a)))$,
cf. (\ref{Poincare}). Consequently, the subalgebras ${\cal A}_c$ and
${\cal B}$ are Poincar\'e-invariant.
\subsubsection{Representing the algebra of observables}
Rieffel induction starts from the input data of diagram
(\ref{diag2}). So far, we have determined the algebra of weak
observables ${\cal A}_c={\cal A}(N_{sk},B_{sk})$ and
the algebra of constraints ${\cal B}={\cal A}(T_{sk},B_{sk})$
of the St\"uckelberg-Kibble
model; note that  ${\cal B}\subset {\cal A}_c$. What is needed
is a representation of these
algebras on some subspace $L$ of a Hilbert space ${\cal H}$. In this
subsection, we give such a representation on a bosonic Fock space
(cf.\ the corresponding procedure for QED in \ci{Landswied}).

For simplicity, in a first step we introduce  a representation for
elements $W({\phi}_{\mu},{\phi}=0) \in {\cal A}(M_{sk},B_{sk})$
only. This will subsequently be generalized to the whole algebra.
We start from the canonical
commutation relations for the smeared annihilation and creation operators
$\hat{a}_{\mu}$, $\hat{a}_{\mu}^*$,
\be
 \hat{a}(f)= \hat{a}_{\mu}(f^{\mu}) =
\int {d^3{\bf k}\over (2\pi)^32k_0} \lbrack{ \hat{a}_0({\bf
k})\overline{f}_0({\bf k}) + \hat{a}_i({\bf k})\overline{f}_i({\bf
k})}\rbrack,
\ee
namely
\be
{\lbrack{ {\hat{a}(f)}, {\hat{a}^*(g)} }\rbrack} = (g,f)_E
:= \int {d^3{\bf k}\over (2\pi)^32k_0}
{g}_{\mu}({\bf k}) {\delta}^{\mu\nu} \overline{f}_{\nu}({\bf
k}). \ll{ccr-euclid}
\ee
For reasons to become clear soon, we have employed the so-called
Fermi trick \ci{CGH} which consists in defining the creation and
annihilation operators of a vector field such that their commutator
is a Euclidean scalar product.
Introducing a vacuum state $|0\rangle$ with the property $
\hat{a}(f)|0\rangle=0$ for all $f$, the creation and annihilation
operators generate a bosonic Fock space $\H_1$ in the usual way.
Mathematically $\H_1$ is, of course, the symmetric Hilbert space
\ci{Guichardet} over  $L^2({\R}^3)\otimes {\C}^4$.

We can now represent the field algebra $\cal A$, and thence its
subalgebras ${\cal A}_c$ and $\cal B$, on $\H_1$ as follows:
\be
 {\pi}(W({\phi}^{\mu}, {\phi}=0)) = e^{\lbrack{
{\hat{a}_{\mu}(\tilde{\phi}_{\mu})} - {\hat{a}_{\mu}(\tilde{\phi}_{\mu})^*}
}\rbrack},
\ee
where $\tilde{\phi}_{\mu}=\left( -\overline{\phi}_0, {\phi}_i \right)$,
and the symbol ${\pi}$ for a representation has been introduced.
The essential point is that the Euclidean commutation relations
(\ref{ccr-euclid}) are able to represent the Minkowski commutators
(\ref{ccrsk}) because of the special definition of $\tilde{\phi}_{\mu}$.

 Now, we present a very economical
notation for symmetric $n$-particle states by introducing `exponential
vectors' \ci{Guichardet}. To this aim, we represent the algebra
${\cal A}_c$ on the dense subset $L_1$ of ${\cal H}_{1}$,
which is the span of all exponential vectors
\bea
 L_1 &=& {\lbrace{\sum_{i=1}^N{\lambda}_{i}e^{{\psi}^{(i)}} \mid
{\lambda}_i\in{\C}, {\psi}^{(i)}\in
L^2({\R}^3)\otimes {\C}^4, N<\infty}\rbrace};
\nn \\
 e^{\psi} &:=& 1\oplus\psi \oplus {1\over \sqrt 2}\psi\otimes\psi
\oplus {1\over \sqrt{3!}}\psi\otimes\psi\otimes\psi \oplus\ldots ,
\ll{exp-vect}
\eea
where the tensor products are understood to be symmetrized.
Note that the prefactors $1\over \sqrt{n!}$ of the $n$-particle
contributions to $e^{\psi}$ have been chosen differently from those of a
Taylor expansion of $e^x$. This allows for a simple form of the
scalar product on $L_1$,
\be
(e^{\psi},e^{\varphi}) = e^{(\psi,\varphi)_E}.\ll{e-scalar-prod}
\ee
A useful remark is now that symmetric $n$-particle states can be
obtained from suitably normalized derivatives of exponential vectors,
\be
{\psi}_1{\otimes}_s ... {\otimes}_s {\psi}_n = {1\over \sqrt{n!}}
{d\over dr_1} ... {d\over dr_n} e^{\sum_ir_i{\psi}_i} |_{r_i = 0}.
\ll{deriv}
\ee

The representation of $W({\phi}_{\mu},0)$ takes a very simple
form on $L_1$. From   (\ref{exp-vect}) we
have
$e^{\hat{a}_{\mu}({\phi}^{\mu})} e^{\psi} =
e^{{(\psi,\phi)_E}}e^{\psi}$,
$e^{\hat{a}_{\mu}({\phi}^{\mu})^*}  e^{\psi} =
e^{(\psi+\phi)}$
and hence\footnote{
To see that this defines a
representation, we check that
$$ {\pi}(W({\phi}_{\mu},0)) {\pi}(W({\varphi}_{\mu},0)) = e^{\lbrack{i{\rm
Im}{\lbrack{ (\overline{\phi}_0,\overline{\varphi}_0)_E +
({\phi}_i,{\varphi}_i)_E}\rbrack}}\rbrack}
{\pi}(W({\phi}_{\mu}+{\varphi}_{\mu},0)),$$
where
$ {\rm Im}{\lbrack{ (\overline{\phi}_0,\overline{\varphi}_0)_E +
({\phi}_i,{\varphi}_i)_E}\rbrack} =
B_{sk}({\varphi}_{\mu},0;{\phi}_{\mu},0)$.}
\be
 {\pi}(W({\phi}_{\mu},0))e^{\psi} = e^{{-1\over
2}(\phi,\phi)_E+(\psi,\tilde{\phi})_E}
e^{(\psi-\tilde{\phi})}.
\ee
%

The construction given above is easily generalized to
the whole algebra ${\cal A}(M_{sk},B_{sk})$ acting on the dense
subspace $ L = L_1 \otimes L_2$
of
$
 {\cal H} ={\cal H}_1\otimes {\cal H}_2$, where ${\cal H}_2$ is the
bosonic Fock space over $L^2(\R^3)$. With
\be
L_2 = {\lbrace{ \sum_i^N {\lambda}_i e^{{\psi}^{(i)}} \mid
{\psi}^{(i)} \in L^2({\R}^3); {\lambda}_i\in\C, N<\infty
}\rbrace}
\ee
the scalar product of vectors in $L$, reads
\be
(e^{{\psi}_{\mu}}\otimes e^{{\psi}},e^{{\chi}_{\mu}}\otimes
e^{{\chi}}) = e^{ ({\psi}_{\mu},{\chi}_{\mu})_E +
({\psi},{\chi})},
\ee
and the action of ${\cal A}(M_{sk},B_{sk})$ (denoted by $\pi$ as well,
with slight abuse of notation) is
\be
{\pi}(W({\phi}_{\mu},{\phi})) e^{{\psi}_{\mu}}\otimes
e^{{\psi}} = e^{{-1\over 2}({\phi}_{\mu},{\phi}_{\mu})_E +
({\psi}_{\mu},\tilde{{\phi}_{\mu}})_E}
e^{{-1\over 2}({\phi},{\phi}) + ({\psi},{\phi})}
e^{{\psi}_{\mu}- \tilde{{\phi}_{\mu}}} \otimes e^{{\psi}-\phi}.
\ll{sk-rep}
\ee
It should be pointed out that $L$ is only stable under finite linear
combinations of the $W$'s (which span a dense subalgebra of $\cal A$),
and not under all elements of $\cal A$.
Hence, strictly speaking, the induction process is performed relative
to the corresponding dense
subalgebras of ${\cal A}_c$ and $\cal B$.
\subsubsection{Constructing the physical one-particle Hilbert space}
With (\ref{sk-rep}), we have specified  the bimodule $L$ for
${\cal A}_c$ and $\cal B$, which in this case is a subspace  of an
`unphysical' Hilbert space ${\cal H}$. Our next step is to construct
the corresponding physical Hilbert space, i.e., to carry out the
discussion following
(\ref{diag2}). In this and the next subsection, we determine the null
space ${\cal N}_{sk}$ for the St\"uckelberg-Kibble model, thereby eventually
obtaining ${\cal H}_{phys}$.

We start from the inner product on elementary vectors in $L$
\be
 (e^{{\psi}_{\mu}}\otimes e^{\psi},
e^{{\chi}_{\mu}}\otimes e^{\chi})_0 = \int_{T_{sk}} [{\cal D}g]
({\pi}(W({\partial}_{\mu}g,-eg))
e^{{\psi}_{\mu}}\otimes e^{\psi},
e^{{\chi}_{\mu}}\otimes e^{\chi}), \ll{pisk}
\ee
which is a natural generalization of (\ref{0-prod}) (and can, at least
heuristically, be derived from an appropriate rigging map defined by a
unitary representation of the gauge group on $\cal H$).
As in \ci{Landswied}, the heuristic path integral (\ref{pisk}) can
be turned into a well-defined integral w.r.t.\ a certain cylindrical
measure on $T_{sk}=G$, but here we shall proceed with the formal flat
measure ${\cal D}g$, and certify that all manipulations below can be
rigorously justified.

 Using the representation  (\ref{sk-rep}) of
${\cal A}(N_{sk},B_{sk})$,  we obtain, with $k_0 = \sqrt{e^2 + {\bf k}^2}$,
and $d\tilde{k} = {d{\bf k}^3\over ({2\pi})^32k_0}$,
\bea
(e^{{\psi}_{\mu}}\otimes e^{\psi}&,&
e^{{\chi}_{\mu}}\otimes e^{\chi})_0\nn \\
&& = e^{\int \tilde{dk}  {-1\over k_0^2}{\lbrack{ (k_i{\psi}_i-
ie\psi)k_0{\psi}_0 +
(k_i\overline{\chi}_i + ie\overline{\chi})k_0\overline{\chi}_0
 }\rbrack}}\nn \\
&&\times e^{ \int \tilde{dk} {\psi}_i{\left({ {\delta}_{ij} -
{k_ik_j\over {\bf
k}^2}}\right)} \overline{\chi}_j + {\left({ {e\over k_0}{\psi}_i + i
{k_i\over k_0}\psi}\right)} {k_ik_j\over {\bf k}^2} \overline{\left({
{e\over k_0}{\chi}_j + i {k_j\over k_0}\chi}\right)}
},\ll{sk-0-prod}
\eea
where we have used
$ {\left({ {\delta}_{ij} - {k_ik_j\over k_0^2}}\right)}
= {\left({ {\delta}_{ij} - {k_ik_j\over {\bf k}^2}}\right)} + {e^2
k_ik_j\over k_0^2 {\bf k}^2}$ to write (\ref{sk-0-prod}) in terms
of projection operators.

To investigate the structure of the null space ${\cal N}_{sk}$, we
derive the $(.,.)_0$-inner product for $n$-particle vectors in
${\cal H}$ from (\ref{sk-0-prod}). For one-particle vectors in
the (unphysical) space ${\cal H}$, we have
\be
{d\over dr} e^{r{\psi}_{\mu}}\otimes e^{r\psi}|_{r=0} =
{\psi}_{\mu} \otimes {\Omega}' + {\Omega}''\otimes\psi,
\ee
where ${\Omega} = {\Omega}''\otimes {\Omega}'$ denotes the vacuum
state in ${\cal H}$. Since such expressions become cumbersome for
higher derivatives, for notational convenience  we define
\be
{\psi}_*^{(1)}\times ...\times {\psi}_*^{(n)}
:= {1\over \sqrt{n!}}{d\over dr_1} ... {d\over dr_n}
e^{\sum_i r_i {\psi}_{\mu}^{(i)}} \otimes e^{\sum_j r_j
{\psi}^{(j)}} {\mid}_{r_i=0}.
\ee

Then, the $(.,.)_0$-inner product on one-particle
vectors in ${\cal H}$ reads
\be
({\psi}_*,{\chi}_*)_0 =
{ \int \tilde{dk} {\psi}_i{\left({ {\delta}_{ij} -
{k_ik_j\over {\bf
k}^2}}\right)} \overline{\chi}_j + {\left({ {e\over k_0}{\psi}_i + i
{k_i\over k_0}\psi}\right)} {k_ik_j\over {\bf k}^2} \overline{\left({
{e\over k_0}{\chi}_j + i {k_j\over k_0}\chi}\right)}}. \ll{one-part-prod}
\ee
Clearly, the two transversal components $P_T{\psi}_* :=
{\left({ {\delta}_{ij} - {k_ik_j\over {\bf k}^2}}\right)} {\psi}_j$
and a linear combination $P_L{\psi}_*$ of the
longitudinal component ${k_ik_j\over{\bf k}^2}{\psi}_j({\bf k})$
with the scalar component ${\psi}({\bf k})$ survive, while the remaining
two components lie in ${\cal N}_{sk}$. To be more precise, we
introduce for ${\psi}_*$ the {\it Bogoliubov-transformed components}
${\psi}_L$, ${\psi}_N$,
\bea
{\psi}_{L,i}({\bf k})  &:=& \cos\theta {k_ik_j{\psi}_j({\bf k})\over
{\bf k}^2} + i\sin\theta {k_i{\psi}({\bf k})\over |{\bf k}|}
,\nn \\
{\psi}_{N,i}({\bf k}) &:=& -\sin\theta {k_ik_j{\psi}_j({\bf k})\over
{\bf k}^2} + i\cos\theta {k_i{\psi}({\bf k})\over |{\bf k}|},
\eea
where
$\cos\theta = {e\over k_0}$, $\sin\theta =
{|{\bf k}|\over k_0}$.
With ${\psi}_L$, ${\psi}_T$ and ${\psi}_N$, the five-component
vector ${\psi}_*^{(i)}$ can be specified as
\be
 {\psi}_*({\bf k}) := \left({P_T{\psi}_{\mu}({\bf k}),
{\psi}_L({\bf k}), {\psi}_N({\bf k}), {\psi}_0({\bf
k})}\right),
\ee
and the projection operator $P_p$ onto the `physical' one-particle
components is given by
\be
(P_p{\psi}_*)({\bf k}) = \left({P_T{\psi}_{\mu}({\bf k}),
{\psi}_L({\bf k}),0,0}\right). \ll{one-part-phys}
\ee
This is exactly what one expects: the five `unphysical' degrees of
freedom have combined into three physical ones in such a way that the
longitudinal component in ${\cal H}$ has mixed with the scalar
component.

\subsubsection{The physical Hilbert space ${\cal H}_{phys}$}
To extend (\ref{one-part-phys}) to $n$-particle states, we
rewrite (\ref{sk-0-prod}), using
$$\exp(\sum_ir_i{\psi}_*^{(i)}) := \exp(\sum_ir_i{\psi}_{\mu}^{(i)})
\otimes \exp(\sum_ir_i{\psi}^{(i)}),$$
\be
(e^{{\psi}_*},e^{{\chi}_*})_0 = (e^{{\psi}_*},\Omega)_0
(\Omega, e^{{\chi}_*})_0 (e^{P_p{\psi}_*},e^{P_p{\chi}_*}). \ll{nullsk}
\ee
Here we have used the remark following  (\ref{one-part-prod}),
which implies that
$$ (\exp(P_p{\psi}_*),\exp(P_p{\chi}_*))_0 =
(\exp(P_p{\psi}_*),\exp(P_p{\chi}_*)).$$
{}From (\ref{nullsk}) we obtain
\bea
{\psi}_*^{(1)}\times ... \times{\psi}_*^{(n)} &=&
{d\over dr_1}...{d\over dr_n} (e^{\sum_ir_i{\psi}_*^{(i)}},\Omega)_0
e^{\sum_ir_i{\psi}_*^{(i)}} |_{r_i=0}\nn \\
&=& \sum_{q=0}^n \sum_{(p_i)_1^q\in {\cal P}_{q,n}}
{\lambda}_{(p_i)_1^q}
(P_p{\psi}_*^{(p_1)}){\times} ... {\times}(P_p{\psi}_*^{(p_q)})
+ \vec{n},\ll{n-part-unphys}
\eea
where ${\cal P}_{q,n}$ contains all sets of $q$ indices
${\lbrace{(p_i)_1^q}\rbrace}$ out of
${\lbrace{1,...,n}\rbrace}$, such that
${\lbrace{(p_i)_1^q}\rbrace}$ $\cup$
${\lbrace{(\hat{p_i})_1^{n-q}}\rbrace}$ $={\lbrace{1,...,n}\rbrace}$.
Here,
\be
{\lambda}_{(p_i)_1^q} =
\sqrt{(q)! (n-q)!\over n!} ({\psi}_*^{(\hat{p}_1)}\times ... \times
{\psi}_*^{(\hat{p}_{n-q})}{\mid}_{p_q(I_{n,q})},\Omega)_0
\ee
are $c$-number coefficients and $\vec{n}$ denotes an element in
${\cal N}_{sk}$.

Vectors of the type (\ref{one-part-phys}) generate a  Hilbert space
of physical one-particle states. The bosonic Fock space over this
one-particle space is evidently ${\cal F}_{phys}:={\cal
S}(L^2({\R}^3)\otimes {\C}^3)$, the symmetric Hilbert space over
$(L^2({\R}^3)\otimes {\C}^3)$. It should be clear from equation
(\ref{n-part-unphys}) that the induced space ${\cal H}_{phys}$ from
the Rieffel induction procedure is naturally isomorphic to this
physical Fock space.\footnote{
Of course, all Hilbert spaces of the same dimension are unitarily
equivalent, but to impose such equivalence one generally has to pick
a basis. We use the term `naturally isomorphic' to indicate that a
unitary equivalence exists which doesn't require the choice of a basis.
{}From the point of view of representation theory, this equivalence
intertwines the actions of appropriate operator algebras,
cf.\ the next subsection.}
To prove this, we define a map $V:L\rightarrow {\cal F}_{phys}$ by linear
extension of $V\exp(\psi_*)=(\exp(\psi_*),\Omega)_0\exp(P_p\psi_*)$.
It follows from an argument similar to the one in section 3.3 of
\ci{Landswied} that this map is well-defined (which is a
nontrivial property, as the basis $\{\exp(\psi_*)\}$ is overcomplete).
Eq.\ (\ref{nullsk}), and the fact that the inner product in
${\cal F}_{phys}$ is just the one in $\cal H$, restricted to the
physcial states, then implies the crucial property
\be
(V\Psi,V\Phi)=(\Psi,\Phi)_0 \ll{crucial}
\ee
for all $\Psi,\Phi\in L$, where the inner product on the l.h.s.\
is evidently the one in ${\cal F}_{phys}$. Hence the null space
${\cal N}_{sk}$ of $(.,.)_0$ is precisely the kernel of $V$,
and the quotient map
$\tilde{V}: L/{\cal N}_{sk}\rightarrow {\cal F}_{phys}$
can be extended to a unitary map (denoted by the same symbol)
$\tilde{V}: {\cal H}_{phys}\rightarrow {\cal F}_{phys}$.
\subsubsection{$n$-point correlation functions and gauge-invariance}
Having specified the physical Hilbert space ${\cal H}_{phys}$, the
next step is to determine the action of ${\pi}_{phys}({\cal A}_c)$.
To this end, we consider  the generating functional ${\omega}_{vac}$ for
vacuum expectation values,
\bea
{\omega}_{vac}({\phi}^{\mu},\phi) &:=&
({\pi}(W({\phi}_{\mu},\phi))\Omega, \Omega)_0 \nn \\
&=& e^{{1\over 2}({\phi}_{\mu},{\phi}_{\mu})_M} e^{-{1\over
2}({\phi},{\phi})} e^{-{1\over
k_0^2}(k_0\overline{\phi}_0(k_{\mu}{\phi}_{\mu} +
ie\phi))}, \ll{vac-exp}
\eea
where $\Omega \in {\cal H}$ is the (unphysical) `vacuum' state.
By construction, only ${\cal A}_c={\cal A}(N_{sk},B_{sk})$ acts on
${\cal H}$ (cf.\ (\ref{nsk})), and for $({\phi}_{\mu},{\phi}) \in N_{sk}$,
$k_0{\phi}_0 = k_i{\phi}_i - ie{\phi}$, we obtain
\bea
{\omega}_{vac}({\phi}^{\mu},\phi)
&=& e^{-{1\over 2}({\phi}_{\mu},P_T{\phi}_{\mu})_E}
e^{-{1\over 2}({e\over k_0}{\phi}_i + i{k_i\over k_0}\phi)
{k_ik_j\over {\bf k}^2} ({e\over k_0}\overline{\phi}_j - i{k_i\over
k_0}\overline{\phi})}\nn \\
&=:& ({\pi}_{phys}(\tilde{W}(P_p{\phi}_*)){\Omega}_{phys},
{\Omega}_{phys})_{phys} \nn \\
&=&
e^{-\half\int d\tilde{k} [\overline{P_T\phi({\bf k})}P_T\phi({\bf k})
+ \overline{\phi}_{L,i}({\bf k}){\phi}_{L,i}({\bf k})]}.
\eea
Here, ${\Omega}_{phys} \in {\cal H}_{phys}$ is the physical
vacuum state; it is just the projection of $\Omega\in L$ onto $L/{\cal
N}_{sk}\, \subset {\cal
H}_{phys}$.

We observe that for $({\phi}_{\mu},\phi)\in
T_{sk}$, ${\pi}(W({\phi}_{\mu},\phi))$ equals the unit operator,
cf.\ (\ref{tsk}). This implies that the gauge group is represented
trivially on ${\cal H}_{phys}$. Moreover, one infers that
${\cal A}_{obs}:= {\pi}({\cal A}_c)\simeq {\cal A}(N_{sk}/T_{sk},B_{c,sk})$,
since the image of a representation of a $C^*$-algebra is isomorphic to
the algebra quotiented by the kernel of the representation.
Now $N_{sk}/T_{sk}\simeq P_p N_{sk}$ as vector spaces (but not as
carrier spaces of actions of the Poincar\'{e} group!), so that,
equally well, ${\cal A}_{obs}\simeq {\cal A}(P_p N_{sk},B_{sk})$.\footnote{
However, the isomorphism between ${\cal A}(P_p N_{sk},B_{sk})$ and
${\cal A}(N_{sk}/T_{sk})$ does not preserve the (automorphic) action
of the Poincar\'{e} group, which, indeed, acts on the latter but not
on the former, cf.\ \ci{CGH}.}
Then, it is clear from section 3.1 that ${\cal A}_{obs}$ is precisely
the Weyl algebra over de Marsden-Weinstein reduced
space (i.e., the physical phase space) of the St\"uckelberg-Kibble model.
Hence it describes three gauge-invariant, massive field components.

Thus  $\tilde{W}(P_p{\phi}_*))$ can be viewed as a Weyl operator in
${\cal A}(P_pN_{sk},B_{sk})$. In particular, the representation of
${\cal A}(P_pN_{sk},B_{sk})$  on exponential vectors
$e^{\psi} \in {\cal H}_{phys} = {\cal S}(L^2({\R}^3)\otimes {\C}^3)$
is given by
\bea
{\pi}_{phys}(\tilde{W}(P_p{\phi}_*))e^{\psi} =
e^{-\half(P_p{\phi}_*,P_p{\phi}_*)_p + (\psi,P_p{\phi}_*)_p}
e^{(\psi - P_p{\phi}_*)} \nn \\
(\psi,P_p{\phi}_*)_p =
\int d\tilde{k} [\overline{P_T\phi({\bf k})}P_T\phi({\bf k})
+ \overline{\phi}_{L,i}({\bf k}){\phi}_{L,i}({\bf k})].
\eea
{}From ${\omega}_{vac}({\phi}_{\mu},{\phi})$, $n$-point correlation
functions can be obtained as multiple derivatives of
$\tilde{W}(P_p{\phi}_*) := e^{i\tilde{A}(f)}$, where
$P_p{\phi}_* = \triangle * f \in L^2({\R}^3)\otimes {\C}^3$.
\bea
&i^n& ({\pi}_{phys}(\tilde{A}(f_1)...\tilde{A}(f_n){\Omega}_{phys},
{\Omega}_{phys})_{phys} =
{d\over dr_1}...{d\over dr_n} {\omega}_{vac}(\sum_ir_i{\phi}_{\mu}^{(i)},
\sum_ir_i{\phi}^{(i)})|_{r_i=0}\nn \\
&=& \sum_{(p_i,q_i)_i^{n\over 2} \in {\cal S}_n} \prod_{i=1}^{n\over 2}
({\pi}_{phys}(\tilde{A}(f_{p_i})\tilde{A}(f_{q_i}){\Omega}_{phys},
{\Omega}_{phys})_{phys}(-1)^{n\over 2} \ll{n-point}
\eea
for $n$ even and zero otherwise. Here, ${\cal S}_n$ denotes the set of all
symmetric partitions of ${\lbrace{1, ...,n}\rbrace}$ into a set of
unordered pairs $(p_i,q_i)$. We conclude from (\ref{n-point})
that the $n$-point correlation functions can be decomposed into
products of $2$-point correlation functions, i.e., Wick's theorem
is satisfied.
The reader should note, however, that this form of Wick's theorem is
satisfied for elements in ${\cal A}(N_{sk},B_{sk})$ only. The crucial
point is that in general, the $(.,.)_0$-inner product preserves the
adjoint for test functions in $N_{sk}$ only. This can be seen by
comparing, e.g.,
${d\over dr_1}{d\over dr_2} ({\pi}(W(\sum_ir_i{\phi}_{\mu}^{(i)},$
$\sum_ir_i{\phi}^{(i)}))\Omega,\Omega)_0 |_{r_i=0}$ with
${d\over dr_1}{d\over dr_2} ({\pi}(W({\phi}_{\mu}^{(1)},$
${\phi}^{(1)}))\Omega, {\pi}(W({\phi}_{\mu}^{(2)},
{\phi}^{(2)}))\Omega)_0 |_{r_i=0}$, cf. (\ref{vac-exp}).

There is an interesting parallel between this restriction of the
Rieffel induced expectation values to ${\cal A}(N_{sk},B_{sk})$ and
the general set-up of the Gupta-Bleuler indefinite metric formalism
as presented in \ci{Strocchi}. In the latter, one starts from an
unphysical Hilbert space ${\cal H}_{GB}$ from which the physical one
is obtained as a quotient ${\cal H}'/{\cal H}''$. Without reviewing
this construction, we note that ${\cal H}$ has to be restricted to a
suitable subspace ${\cal H}' \subset {\cal H}_{GB}$ before quotiening
by a null space ${\cal H}'$. Obviously, in our  setting, a
similar restriction is needed on the level of the algebra,
${\cal A}(N_{sk},B_{sk})\subset {\cal A}(M_{sk},B_{sk})$.
This restriction emerges in a systematic way, for as we pointed out
before, the subalgebra in question is the commutant of the algebra
generated by the constraints (i.e., by the gauge group).

This observation is closely related to the result of Narnhofer and
Thirring \ci{Narnhofer} that covariant formulations without indefinite
inner metric are possible as long as the representation on the physical
Hilbert space is restricted to a certain subalgebra of weak observables.
In the example of Narnhofer and Thirring, non-regular states have
to be introduced. This can be avoided in the Rieffel induction  setting,
cf. \ci{Landswied,UAW} for further details.

\subsubsection{Positivity of the Hamiltonian and action of the
Poincar\'e group}
On the algebra of weak observables of the St\"uckelberg-Kibble model
${\cal A}(N_{sk},B_{sk})$, the time evolution is given
as an automorphism group ${\tau}_t$,
\be
{\tau}_t[W({\phi}_{\mu},\phi)] = W(e^{it\sqrt{D+e^2}}{\phi}_{\mu},
e^{it\sqrt{D+e^2}}\phi ),
\ee
where $(D{\phi})_{\mu} = (-{\triangle}{\phi}_0, -{\triangle}{\phi}_1,
-{\triangle}{\phi}_2, -{\triangle}{\phi}_3)$.
We want to construct the Hamiltonian $H$, corresponding to ${\tau}_t$
on ${\cal H}$. $H$ is a
representation-dependent operator, implementing the time evolution
${\tau}_t$ in the representation ${\pi}$ by
\be
e^{itH}{\pi}(W({\phi}_{\mu},\phi)) e^{-itH} =
{\pi}({\tau}_t[W({\phi}_{\mu},\phi)]).
\ee
Comparing this with the explicit form of the representation in terms of
annihilation and creation operators $\hat{a}_{\mu}^*$, $\hat{a}_{\mu}$
for the vector field and $\hat{b}^*$, $\hat{b}$ for the
scalar field, we obtain
\be
H = - \int d\tilde{k} \sqrt{{\bf k}^2+e^2} \hat{a}_{\mu}^*({\bf k})
g^{\mu\nu} \hat{a}_{\nu}^({\bf k}) + \int d\tilde{k} \sqrt{{\bf
k}^2+e^2} \hat{b}^*({\bf k}) \hat{b}({\bf k}).
\ee
Regarded as an operator on $\cal H$ (with its Hilbert space inner product),
this Hamiltonian clearly has the entire real axis as its spectrum. However,
it is easy to see that
\be
({\Psi}, H {\Psi})_0 \geq 0
\ee
for all $\Psi \in {\cal H}$.
The point is that arbitrary (normalized) components of the physical
one-particle state space,
${\left({ {\delta}_{ij} - {k_ik_j\over {\bf k}^2}}\right)}{\psi}_j$ and
${k_i\over {\bf k}}{\psi}_i \cos\theta + i\psi\sin\theta$
pick up (the same) positive energy contributions. For multi-particle
states, this holds true due to their decomposition into such
components. The elements of ${\cal H}$ carrying the negative
energy spectrum have ended up in the null space. Hence the induced
Hamiltonian $H_{phys}$ on ${\cal H}_{phys}$ is positive.

Finally, we note that ${\cal H}_{phys}$ carries a massive
representation of the Poincar\'e group ${\cal P}$. Indeed, ${\omega}_{vac}$
is Poincar\'e invariant on $N_{sk}$ and hence \ci{Haag,BW} there exists a
Poincar\'e invariant vacuum state ${\Omega}_{phys}\in {\cal H}_{phys}$
and a representation $U_p$ of the Poincar\'e group, such that
\be
U_p(\Lambda,a){\pi}_{phys}(W({\phi}_{\mu},\phi)){\Omega}_{phys} =
{\pi}_{phys}(W({\gamma}_{\Lambda,a}({\phi}_{\mu},\phi))){\Omega}_{phys}
\ee
for all $({\phi}_{\mu},\phi)\in N_{sk}$. It is easily shown that $H_{phys}$
is the generator of the time-translation part of the representation
thus defined. Since the spectrum of the Hamiltonian $H_{phys}$
shows a mass gap, we are dealing with a massive
representation ($m^2 = e^2$) of the Poincar\'e group, i.e.,
the three components of the vector $P_p{\psi}_*^{(i)}$ transform
as a massive one-particle state under the action of the little group
$SO(3)$ \ci{Barut}.

We conclude that ${\cal H}_{phys}$ has the main properties required
by a physical Hilbert space: it transforms trivially under the gauge
group, satisfies the positive spectrum condition and carries a
unitary representation of the Poincar\'e group.

\section{Conclusion}
The quantization proposal employed in this paper provides a detailed
scheme for imposing
constraints on gauge quantum field theories. As explained in Chapter 2,
the main tool of this proposal is the Rieffel induction procedure,
which provides a systematic scheme for the construction of
representations of $C^*$-algebras. It
may be viewed as the quantum counterpart of the  symplectic reduction
technique; as we have shown, this is particulary obvious for Weyl
$C^*$-algebras. This leads to a new
quantization method for gauge field theories.

In the present work, we have applied this method to the
St\"uckelberg-Kibble model. To this end, we have defined  a field
algebra ${\cal A}$ corresponding to the field content of the
Lagrangian,  and an algebra of constraints ${\cal B}$ corresponding
to the gauge group acting on ${\cal A}$. Also, we have specified
a representation ${\pi}$ of subalgebras of ${\cal A}$ on a
(unphysical) Hilbert space ${\cal H}$. From these input data, we have
constructed a representation of the physical, gauge-invariant
fields on a new Hilbert space ${\cal H}_{phys}$.

The construction of ${\cal H}_{phys}$ shows some parallels to the
Gupta-Bleuler indefinite metric formalism. In both settings, a
degenerate inner product is defined on a (unphysical) Hilbert
space ${\cal H}$, and ${\cal H}_{phys}$ is constructed by
quotiening ${\cal H}$ by a null space with respect to this degenerate
inner product. Yet, there are important differences. In contrast to the
indefinite metric inner product ${\langle{.,.}\rangle}$, defined on
${\cal H}_{GB}$ in the Gupta-Bleuler formalism, the $(.,.)_0$-inner
product is positive semidefinite. More importantly, it is
a conceptual advantage of our quantization method that
$(.,.)_0$ is derived from first principles (namely from the
requirement to impose quantum constraints by a quantized version
of the classical phase space reduction method), whereas the
Gupta-Bleuler formalism takes ${\langle{.,.}\rangle}$ as
starting point without further justification. A similar comment
applies to the BRST technique: although a classical analogue of
this procedure exists, the quantum BRST procedure is {\em not} in
any satisfactory sense the quantization of the classical scheme.

Another remarkable difference between both formalisms
is that the Gupta-Bleuler formalism restricts the unphysical Hilbert
space before forming the quotient while the proposal of \ci{Landsman}
restricts itself to  a representation of the subalgebra
${\cal A}_c$ of weak observables on ${\cal H}$, before quotiening
by the appropriate null space.
As a consequence, the $(.,.)_0$-inner product preserves
the adjoint for elements in ${\cal A}_c$ only. It remains to be
seen how far this feature alters applications of usual perturbative
techniques in more complicated models.

Most of our effort in Chapter 3 has gone into characterizing the
particular features of the physical state space ${\cal H}_{phys}$.
By construction, ${\cal H}_{phys}$ carries a trivial representation
of the gauge group. Also, the states are physical in the sense that
they obey a positive spectrum condition and that they carry a
massive representation of the Poincar\'e group. Since the
St\"uckelberg-Kibble model has been widely used in investigations
of the Higgs mechanism, we emphasize again the result obtained
for the one-particle subspace in ${\cal H}_{phys}$. The point is
that in our proposal, the particular construction method of
${\cal H}_{phys}$ allows one to trace back how the (unphysical) components
of ${\cal H}$ end up in the physical Hilbert space. In the present
case, we have shown that the longitudinal physical one-particle
component arises from a particular Bogoliubov-transformation of the
unphysical longitudinal and the scalar component. As expected from
general considerations, two of the five components in ${\cal H}$ have
ended up in the one-particle null space.

We conclude our discussion of the Rieffel induction procedure by
pointing out that our presentation has focused on a
particular way of applying Rieffel induction to gauge quantum field
theories. Conceptually, the scheme is much wider. It remains
to be seen how far other choices for the inner product $(.,.)_0$ and
the unphysical Hilbert space ${\cal H}$ allow for other realisations
of the physical observables of gauge field theories.

 \end{document}